\documentclass[preprint]{elsarticle}

\usepackage{lineno}
\modulolinenumbers[1]

\usepackage{soul}

\usepackage{lipsum}
\usepackage{rotating}
\graphicspath{{figures/}}
\usepackage{pstricks, pst-node, psfrag}
\usepackage{amssymb,amsmath}
\usepackage{mathtools}
\usepackage{verbatim,enumerate}
\usepackage{rotating, lscape}
\usepackage{setspace}
\usepackage[hang, flushmargin]{footmisc}
\usepackage{subfig}
\usepackage{caption}
\usepackage{cancel}
\usepackage{float}
\usepackage{ragged2e}
\usepackage{centernot}
\usepackage{tikz}
\usepackage{tikzsymbols}
\usetikzlibrary{shapes,arrows,positioning,plotmarks}
\usepackage{textcomp} 
\usepackage{gensymb} 

\tikzstyle{endpt} = [rectangle, draw, fill=red!20,
    text width=9.8em, text centered, rounded corners, minimum height=4em]
\tikzstyle{block} = [rectangle, draw, top color=white, bottom color=blue!20,
    text width=9.8em, text centered, rounded corners, minimum height=4em]
\tikzstyle{line} = [draw, -latex', very thick]

\usepackage{listings}
\usepackage{color} 
\definecolor{mygreen}{RGB}{28,172,0} 
\definecolor{mylilas}{RGB}{170,55,241}
\definecolor{mygray}{rgb}{0.5,0.5,0.5}
\definecolor{mycyan}{RGB}{0,255,255}
\definecolor{magenta}{rgb}{1,0,1}

\definecolor{backgreen}{rgb}{0.00, 0.169, 0.212}
\definecolor{textgray}{rgb}{0.514, 0.580, 0.589}

\usepackage[T1]{fontenc}
\newcommand*\patchAmsMathEnvironmentForLineno[1]{%
  \expandafter\let\csname old#1\expandafter\endcsname\csname #1\endcsname
  \expandafter\let\csname oldend#1\expandafter\endcsname\csname end#1\endcsname
  \renewenvironment{#1}%
     {\linenomath\csname old#1\endcsname}%
     {\csname oldend#1\endcsname\endlinenomath}}%
\newcommand*\patchBothAmsMathEnvironmentsForLineno[1]{%
  \patchAmsMathEnvironmentForLineno{#1}%
  \patchAmsMathEnvironmentForLineno{#1*}}%
\AtBeginDocument{%
\patchBothAmsMathEnvironmentsForLineno{equation}%
\patchBothAmsMathEnvironmentsForLineno{align}%
\patchBothAmsMathEnvironmentsForLineno{flalign}%
\patchBothAmsMathEnvironmentsForLineno{alignat}%
\patchBothAmsMathEnvironmentsForLineno{gather}%
\patchBothAmsMathEnvironmentsForLineno{multline}%
}

\newcommand{\cG}{\mathcal{G}}

\newcommand{\cN}{\mathcal{N}}

\newcommand{\cP}{\mathcal{P}}

\newcommand{\cU}{\mathcal{U}}
\newcommand{\cW}{\mathcal{W}}

\newcommand{\dt}{\Delta t}

\newcommand*\ared{}
\newcommand*\ablue{}
\newcommand*\aviolet{}
\newcommand*\agray{}
\newcommand*\aend{}
\newcommand*{\rred}[1]{}
\newcommand*{\rblue}[1]{}
\newcommand*{\rviolet}[1]{}

\newcommand\marksymbol[4]{\tikz[#2,scale=#4,fill=#3]\pgfuseplotmark{#1} ;}

\renewcommand{\vec}[1]{\boldsymbol{#1}}
\renewcommand{\bar}{\overline}
\renewcommand{\tilde}{\widetilde}
\renewcommand{\hat}{\widehat}

\newcommand{\defeq}{\coloneqq}

\DeclareMathOperator\diag{diag}

\journal{Advances in Water Resources}

\begin{document}

\begin{frontmatter}

\title{Numerical Equivalence Between SPH and Probabilistic Mass Transfer Methods for Lagrangian Simulation of Dispersion\tnoteref{mytitlenote}}
\tnotetext[mytitlenote]{This work was partially supported by the US Army Research Office under Contract/Grant number W911NF-18-1-0338; the National Science Foundation under awards EAR-1417145 and DMS-1614586; and the Spanish Ministry of Economy and Competitiveness through project WE-NEED, PCIN-2015-248.}

\author{Guillem Sole-Mari\fnref{barca,h2ogeo}}
\ead{guillem.sole.mari@upc.edu}
\address{Universitat Polit\`ecnica de Catalunya\\ C/ Jordi Girona 1-3\\ 08034 Barcelona, Spain}
\author{Michael J. Schmidt\fnref{ams,hydro}}
\ead{mschmidt1@mines.edu}
\author{Stephen D. Pankavich\fnref{ams}}
\ead{pankavic@mines.edu}
\author{David A. Benson\fnref{hydro}}
\ead{dbenson@mines.edu}
\address{Colorado School of Mines\\ 1500 Illinois St.\\ Golden, CO 80401}
\fntext[barca]{Department of Civil and Environmental Engineering (DECA), Universitat Polit\`ecnica de Catalunya, Barcelona, Spain}
\fntext[h2ogeo]{Hydrogeology Group (GHS), UPC-CSIC, Barcelona, Spain}
\fntext[ams]{Department of Applied Mathematics and Statistics, Colorado School of Mines, Golden, CO, 80401, USA}
\fntext[hydro]{Hydrologic Science and Engineering Program, Department of Geology and Geological Engineering, Colorado School of Mines, Golden, CO, 80401, USA}

\begin{abstract}

Several Lagrangian methodologies have been proposed in recent years to simulate advection-dispersion of solutes in fluids as a mass exchange between numerical particles carrying the fluid. In this paper, we unify these methodologies, showing that mass transfer particle tracking (MTPT) algorithms can be framed within the context of smoothed particle hydrodynamics (SPH), provided the choice of a Gaussian smoothing kernel whose bandwidth depends on the dispersion and the time discretization. Numerical simulations are performed for a simple dispersion problem, and they are compared to an analytical solution. Based on the results, we advocate for the use of a kernel bandwidth of the size of the characteristic dispersion length $\ell=\sqrt{2D\Delta t}$, at least given a ``dense enough'' distribution of particles, for in this case the mass transfer operation is not just an approximation, but in fact the exact solution, of the solute's displacement by dispersion in a time step.

\end{abstract}


\end{frontmatter}


\section{Introduction} 
\label{sec:introduction}

In recent years, a number of Lagrangian numerical schemes have been proposed to simulate advection-dispersion processes in fluids. Some of these approaches rely exclusively on traditional random walks to simulate dispersion \cite{benson2008,Paster_WRR,Paster_JCP,Benson_AWR_2016,Bolster2016,dong_awr,Ding_monod,Ding_WRR,Bolster_mass,schmidt2017,guillem2017kde,sole-mari2018}, whereas a second class represents dispersion through mass transfer between particles that carry a given amount of fluid \cite{herrera_2009,herrera_2013,Benson_arbitrary,mass_trans_acc}.
Other authors have hybridized random walks with mass transfer \cite{Engdahl_WRR,herrera_2017} in an approach that allows partitioning of total dispersion between mixing (simulated by mass transfer) and non-mixed spreading (simulated via random walks).
Mass-transfer algorithms can be further subdivided into two groups.
The first group \cite{herrera_2009,herrera_2013} derives the mass exchange rates from the well-established smoothed particle hydrodynamics (SPH) method \cite{Gingold_originalSPH}, which, besides solute transport, has been used in a variety of applications \cite{Monaghan_SPHappl} such as astrophysics, fluid dynamics, and solid mechanics.
A second group of approaches, often referred to as mass transfer particle tracking (MTPT) algorithms \cite{Benson_arbitrary,mass_trans_acc}, derive the mass-exchange rate from stochastic rules governing the co-location probability of particles moving via dispersion.
To date, a relationship between these two methodologies for mass transfer has not been established.
In this paper we analytically derive the connection between the SPH and MTPT conventions and show that, for specific kernel choices and provided that equivalent normalization and averaging conventions are used, the SPH and MTPT approaches are numerically equivalent.
Additionally, for the fixed choice of a Gaussian kernel, we investigate the effect of differing bandwidth choices on deviations from the analytical, well-mixed solution.


\section{The link between SPH and MTPT} 
\label{sec:link}

The SPH approach to approximating dispersion can be summarized by following \cite{herrera_2009,herrera_2013}.
Therein, the following equation describes the time evolution of the concentration, $C_i(t)$, carried by a numerical particle labeled $i=1,..,N$, \ablue assuming that all particles contain the same amount of fluid\aend:
\begin{equation}\label{s1e1}
    \frac{d C_i}{d t}=2\sum^N_{j=1}{\frac{{\hat{D}}_{ij}}{{\widehat{\rho }}_{ij}}\left(C_i-C_j\right)F\left({\vec{X}}_i-{\vec{X}}_j; h\right)}.
\end{equation}
Here, $N$ is the number of particles, $\vec X_i$ is the position of particle $i$, and $F\left(\vec{r}; h\right)$ is a radial function satisfying
\begin{equation}\label{s1e2}
    \vec{r}F\left(\vec{r}; h\right)=\nabla W\left(\vec{r}; h\right),
\end{equation}
with $W$ representing a radially symmetric, translation-invariant kernel with bandwidth $h$.
Additionally, ${\hat{D}}_{ij}$ is the effective dispersion coefficient that, in the isotropic but spatially variable case, reduces to
\begin{equation}\label{s1e3}
    {\hat{D}}_{ij} \defeq g\left(D({\vec{X}}_i),D({\vec{X}}_j)\right),
\end{equation}
where $g$ is an averaging function (e.g., arithmetic or harmonic average).
The quantity ${\widehat{\rho }}_{ij}$, defined by
\begin{align}
    \hat{\rho }_{ij}  &\defeq g\left(\rho_i, \rho_j \right),\label{rhoij}\\
    \rho \left(\vec X; h\right) &\defeq \sum_{k = 1}^{N}W\left( {\vec{X}}- {\vec{X}}_k ; h\right),\quad\rho_q\defeq\rho \left(\vec X_q;h\right),\quad q= i, j,\label{rhok}
\end{align}
is an average of the particle densities estimated at ${\vec{X}}_i$ and ${\vec{X}}_j$. 
A popular choice for $g$, in this case, is the arithmetic average.
Note that we make explicit the previously suppressed dependence of $\hat \rho_{ij}$ on the locations of the particles, $\vec X_i,\ i = 1, \dots, N$, and the parameter $h$, which represents the bandwidth of the kernel function $W$.

In the specific case that $W\left(\vec{r}; h\right)$ is a Gaussian kernel with the form
\begin{equation}\label{s1e5}
    W\left(\vec{r}; h\right)={\left(2\pi h^2\right)}^{-d / 2}\ \exp\left({-\frac{\left\vert {\vec{r}} \right\vert^{2}}{2h^2}}\right),
\end{equation}
where $d$ is the number of spatial dimensions, we have
\begin{equation}\label{s1e6}
    F\left(\vec{r}; h\right)=-\frac{1}{h^2}W\left(\vec{r}; h\right).
\end{equation}
Substituting \eqref{s1e6} into \eqref{s1e1}, integrating the expression (first-order explicit), and then rearranging, we arrive at
\begin{equation}\label{s1e7}
    {\ablue C}_i\left(t+\Delta t\right)={\ablue C}_i\left(t\right)+\sum^N_{j=1}{{\beta }_{ij}\cW_{ij}\left({\ablue C}_j\left(t\right)-{\ablue C}_i\left(t\right)\right)},
\end{equation}
in which we define
\begin{align}
    {\beta }_{ij}\left(h\right) &\defeq \frac{\ell_{ij}^2}{h^2},\quad \ell_{ij}\defeq \sqrt{2{\hat{D}}_{ij}\Delta t},\label{beta_ij}\\
    \cW_{ij}\left(h\right) &\defeq \frac{W\left({\vec{X}}_i-{\vec{X}}_j; h\right)}{{\widehat{\rho }}_{ij}\left(h\right)}.\label{normed_W}
\end{align}
Here, we once again denote the dependence of $\beta$ and $\cW$ on $h$ because, for a different kernel bandwidth choice, these quantities will be altered correspondingly. Note that $\ell_{ij}$ in \eqref{beta_ij} is equal to the characteristic distance of the average dispersion of particles $i$ and $j$ in a time step $\Delta t$.

We now consider, for the sake of comparison, the MTPT algorithm originally formulated by \emph{Benson and Bolster} \cite{Benson_arbitrary}, further discussed in \cite{mass_trans_acc}, and given by
\begin{equation}\label{arb_rxn_MT}
    {\ablue C}_i\left(t+\Delta t\right)={\ablue C}_i\left(t\right) + \frac{1}{2}\sum^N_{j=1} \cP_{ij}\left({\ablue C}_j\left(t\right)-{\ablue C}_i\left(t\right)\right),
\end{equation}
where $\cP_{ij}$ is the probabilistic weighting function for a mass transfer from particle $j$ to particle $i$, with the form
\begin{equation}\label{normP}
    \cP_{ij} = \frac{P(\vec X_i - \vec X_j)}{\tilde \rho_{ij}}.
\end{equation}
Here, the function $P$ is the probability density for the co-location of particles $i$ and $j$, moving via dispersion,
\begin{align}\label{coll_prob}
    P(\vec X_i - \vec X_j; D, \dt) &= \left(4 \pi (D_i + D_j) \dt\right)^{-d / 2} \exp \left[-\frac{\vert \vec X_i - \vec X_j \vert^2}{4 (D_i + D_j) \dt}\right]\\
    &\equiv W\left(\vec X_i - \vec X_j ; \sqrt{2(D_i + D_j) \dt}\right) \nonumber\\
    &= W\left(\vec X_i - \vec X_j ; \sqrt{4 \hat D_{ij} \dt}\right)  \nonumber\\
    &=W\left(\vec X_i - \vec X_j ; \sqrt{2}\ell_{ij}\right), \nonumber
\end{align}
where $D_k \defeq D(\vec X_k)$, and $\tilde \rho_{ij}$ is a normalizing factor that has classically been chosen to be $\rho_j$, as in \eqref{rhok}, with $h=\sqrt{2}\ell_{ij}$,
in order to make the matrix $\vec{\cP}$ (with $i,j^{\text{th}}$ entry $\cP_{ij}$) a left stochastic matrix {\ared (i.e., a matrix where all columns sum to 1)}. 
\ablue However, this does not guarantee that $\cP_{ij}=\cP_{ji}$. Hence, the concentration increase (or decrease) at particle $i$ due to its interaction with particle $j$ by \eqref{arb_rxn_MT} may not match the decrease (or increase) at particle $j$ due to interaction with particle $i$. As a consequence of this asymmetry, normalization by $\tilde\rho_{ij}=\rho_j$ may not impose exact mass conservation. Also, in the original paper \cite{Benson_arbitrary}, equation \eqref{arb_rxn_MT} is formulated in terms of solute masses instead of concentrations, which are, in this case, interchangeable since all particles carry an equal amount of fluid.\aend

Comparing equations \eqref{s1e7} and \eqref{arb_rxn_MT}, it is evident that the co-location probability-based mass exchange algorithm of \emph{Benson and Bolster} \cite{Benson_arbitrary} is numerically equivalent to the SPH formalism for $\beta_{ij}=1/2$ for all $i,j=1,...,N$, with the standard deviation associated with particle co-location by dispersion used for the bandwidth of $W$ in \eqref{s1e5}.
Note that, according to \eqref{beta_ij} and \eqref{normed_W}, imposing a constant value for ${\beta }_{ij}$ implies that the kernel bandwidth $h$ will change with the positions of particles $i$ and $j$ for spatially-variable dispersion and will depend on $\Delta t$, as can be seen from \eqref{coll_prob}.

Expressions \eqref{s1e7} and \eqref{arb_rxn_MT} can be written in a general matrix-vector form as
\begin{equation}\label{s1e10}
    \vec{{\ablue C}}\left(t+\Delta t\right)=\vec{A}\left(t\right)\vec{{\ablue C}}\left(t\right),
\end{equation}
where $\vec{{\ablue C}}_i \defeq {\ablue C}_i$, and
\begin{equation}\label{s1e11}
    \vec{A} \defeq \vec{I}+\left[\vec{\beta}\circ \vec{\cW}-\diag\left(\left[\vec{\beta}\circ \vec{\cW}\right]\vec{1}\right)\right].
\end{equation}
Above, $\vec{I}$ is the $N \times N$ identity matrix, $\vec{1}$ is an $N \times 1$ vector of ones, $\circ $ denotes the entrywise, or Hadamard, product, $\diag(\vec x)$ is a square matrix with the entries of $\vec x$ on its main diagonal, and the $i,j^{\text{th}}$ entries of the matrices $\vec \beta$ and $\vec \cW$ are $\beta_{ij}$ and $\cW_{ij}$, respectively.
Note that, as mentioned above and elsewhere (see \cite{mass_trans_acc}), choosing $\tilde \rho_{ij}$ to be $\rho_j$ in \eqref{normP} ensures that $\vec \cP$ {\agray (denoted as $\vec \cW$ in \eqref{s1e11})} is a left stochastic matrix {\ablue(but not necessarily symmetric)}\rred{ (i.e., the columns sum to 1)}.
On the other hand, we note from \eqref{s1e11} that if $\vec{\cW}$ is symmetric, then $\vec{A}$ is also symmetric with rows and columns that sum to 1, guaranteeing conservation of mass. Thus, a better normalization approach is to choose $\tilde \rho_{ij}$ to be $\hat \rho_{ij}$, as in \eqref{rhoij}, resulting in symmetric $\vec \cW$ and mass-conserving $\vec A$.

\emph{Schmidt et al.} \cite{mass_trans_acc,Schmidt_fluid_solid} also present a discretized Green's function approach to simulating dispersion by mass transfer.
For a time step $\Delta t$, this algorithm is described as:
\begin{equation}\label{Gfunc}
    \vec{{\ablue C}}\left(t+\Delta t\right)=\vec{\cG}\left(t\right)\vec{{\ablue C}}\left(t\right),
\end{equation}
with
\begin{align}\label{s1e13}
    \cG_{ij} &\defeq \frac{W\left({\vec{X}}_i-{\vec{X}}_j;\ell_{ij}\right)}{\tilde \rho_{ij}},
\end{align}
where, once again, $\tilde \rho_{ij}$ is traditionally defined to be $\rho_j$ as in \eqref{rhok}.
We see that the matrix $\vec{\cG}$ is nearly identical to $\vec \cW$ for $h=\ell_{ij}$ (and to $\vec \cP$ with twice the square bandwidth), the only difference being the choice of non-symmetric normalization using $\tilde \rho_{ij}=\rho_j$.

{\ablue We note that, for a sufficiently large $N$, $\sum_{j=1}^N\cG_{ij}\approx\int\rho\left(\vec{x}\right)\frac{W\left(\vec{X}_i-\vec{x}\right)}{\rho\left(\vec{x}\right)}\mathrm{d}\vec{x}=1$, which implies} $\diag\left(\vec{\cG}\vec{1}\right)\approx \vec{I}$. Hence, {\ablue knowing that, for $h=\ell_{ij}$, $\vec{\cW}\approx\vec{\cG}$}, we see that the discretized Green's function algorithm \eqref{Gfunc} is also nearly identical to the SPH and particle co-location expression given in \eqref{s1e10} and \eqref{s1e11}, under the constraint that ${\beta }_{ij}=1$ for all $i,j$. Hereafter, for simplicity, we refer to any matrix $\vec \beta$ with all-equal entries as a scalar $\beta$.

Thus, we have unified the \rblue{seemingly}{\ablue previously} divergent approaches to simulating dispersion that are employed by the SPH and MTPT algorithms.
Namely, to frame things in the SPH context, the MTPT algorithms hold the mass-transfer scaling parameter $\beta$ constant {\agray ($1/2$ or $1$)} and adapt the kernel, itself, to the magnitude of dispersion over a time step.
This is in contrast to the traditional SPH approach, {\ablue where the kernel bandwidth is independent from the dispersion magnitude, and often set to contain a prescribed ``number of neighbors'', either locally or on average \cite{Tartakovsky2016}. The kernel is then} scaled in amplitude by the parameter $\beta_{ij}$ to capture the magnitude of the dispersion action.

Having established the link, through the parameter $\boldsymbol \beta$ in \eqref{s1e11} (alternatively viewed as the choice of kernel bandwidth $h$), between the SPH and MTPT formalisms for simulating dispersion in a Lagrangian context, we next consider the implications of varying this parameter.
In the following section, we conduct some numerical experiments to consider these effects.


\section{Numerical investigations} 
\label{sec:numerical}

\begin{figure}[t]%
    \centering
    \includegraphics[width=1\textwidth]{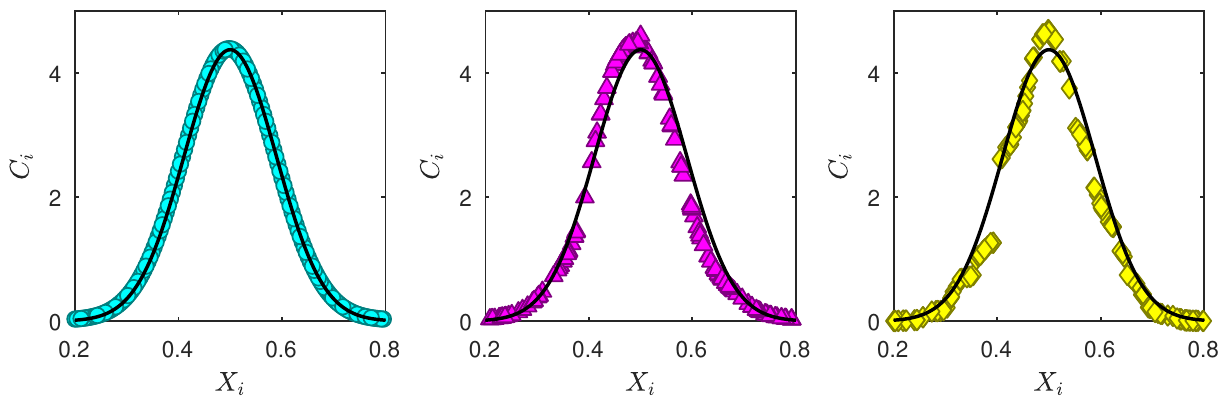}
    \caption{Concentrations at $t=4$ for example realizations with \protect \marksymbol{*}{blue!50!green}{mycyan}{1.2} evenly-spaced, \protect \marksymbol{triangle*}{black!50!magenta}{magenta}{1.5} randomly-spaced, and \protect \marksymbol{diamond*}{black!50!yellow}{yellow}{1.5} random-walking particles. The black line is the analytical solution. For these simulations $N=255$, $\Delta t=0.01$ and $h=h_*$ in each case. The initial condition is a Dirac delta positioned at the center of the domain. Note that deviations from the analytical solution are caused by irregular, and possibly wide, particle spacings, due to low particle numbers in the non-equally-spaced cases.}
    \label{fig:example}
\end{figure}

\begin{figure}[t]%
    \centering
    \includegraphics[width=1\textwidth]{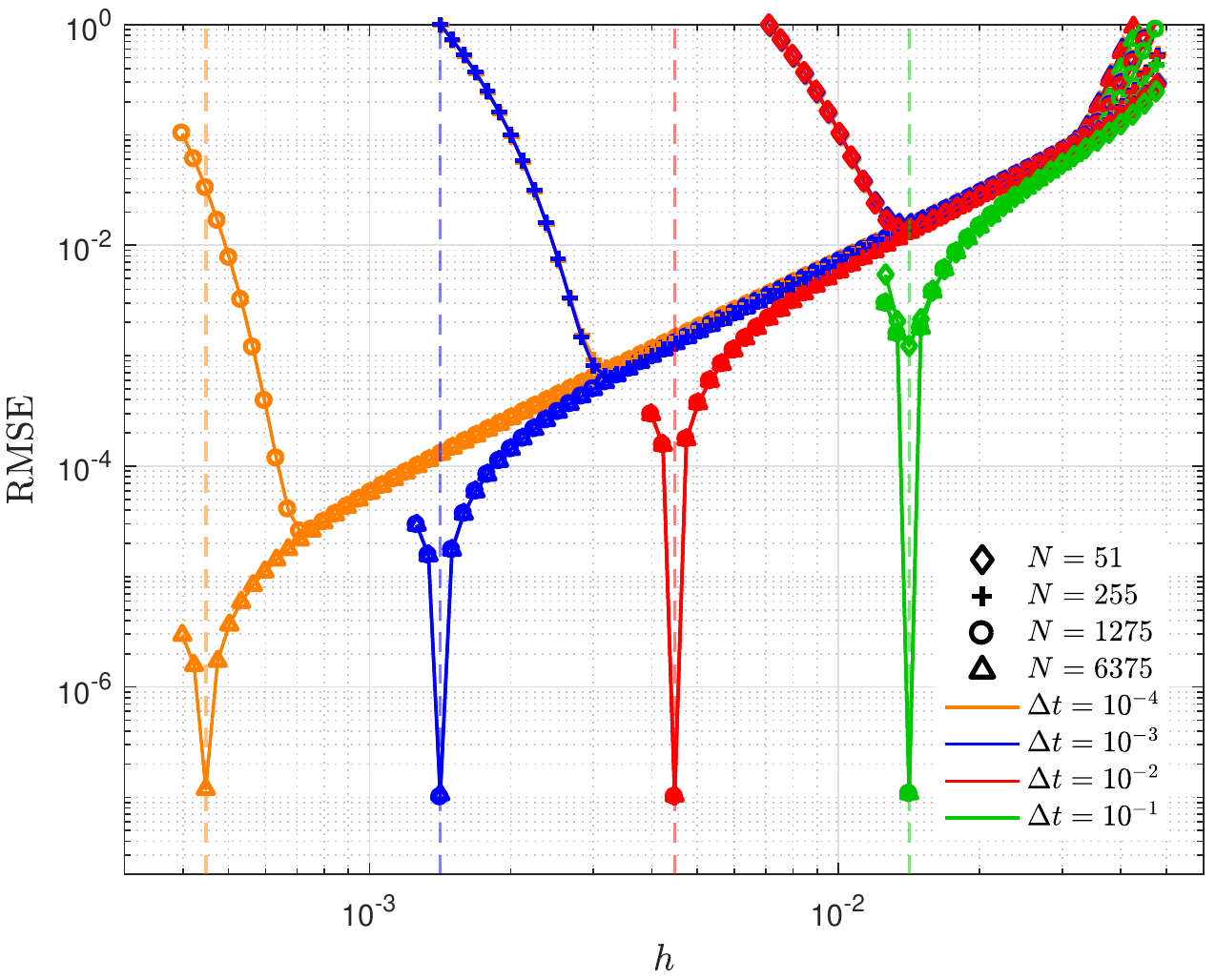}
    \caption{Numerical results for evenly-spaced, stationary particles. $\mathrm{RMSE}$ \eqref{RMSE}, as a function of the kernel bandwidth $h$, is given for different combinations of $N$ and $\Delta t$. The dashed, semitransparent vertical lines indicate the values of $h$ that correspond to $\beta=1$ ($h = \ell = \sqrt{2D\dt}$) for a given value of $\Delta t$.}
    \label{fig:RMSE_equi}
\end{figure}

\begin{figure}[t]%
    \centering
    \includegraphics[width=1\textwidth]{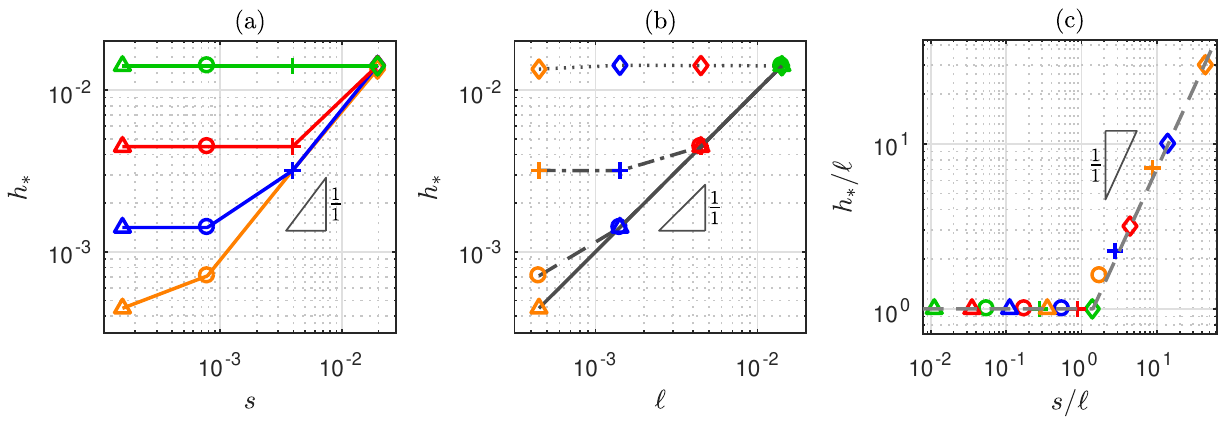}
    \caption{Numerical results for evenly-spaced, stationary particles. $\left( \mathrm a \right)$ Bandwidth $h_*$ associated with the minimum $\mathrm{RMSE}$ plotted against the particle spacing $s=L/N$, for different $\Delta t$ values (see color legend on Figure \ref{fig:RMSE_equi}). $\left( \mathrm b \right)$ Bandwidth $h_*$ associated with the minimum $\mathrm{RMSE}$ plotted against the dispersion distance $\ell=\sqrt{2D\Delta t}$ given different $N$ values (see marker legend on Figure \ref{fig:RMSE_equi}). $\left( \mathrm c \right)$ Lowest-error bandwidth $h_*$ against particle spacing $s=L/N$, both normalized by the dispersion distance $\ell=\sqrt{2D\Delta t}$.}
    \label{fig:hopt_equi}
\end{figure}

\begin{figure}[t]%
    \centering
    \includegraphics[width=1\textwidth]{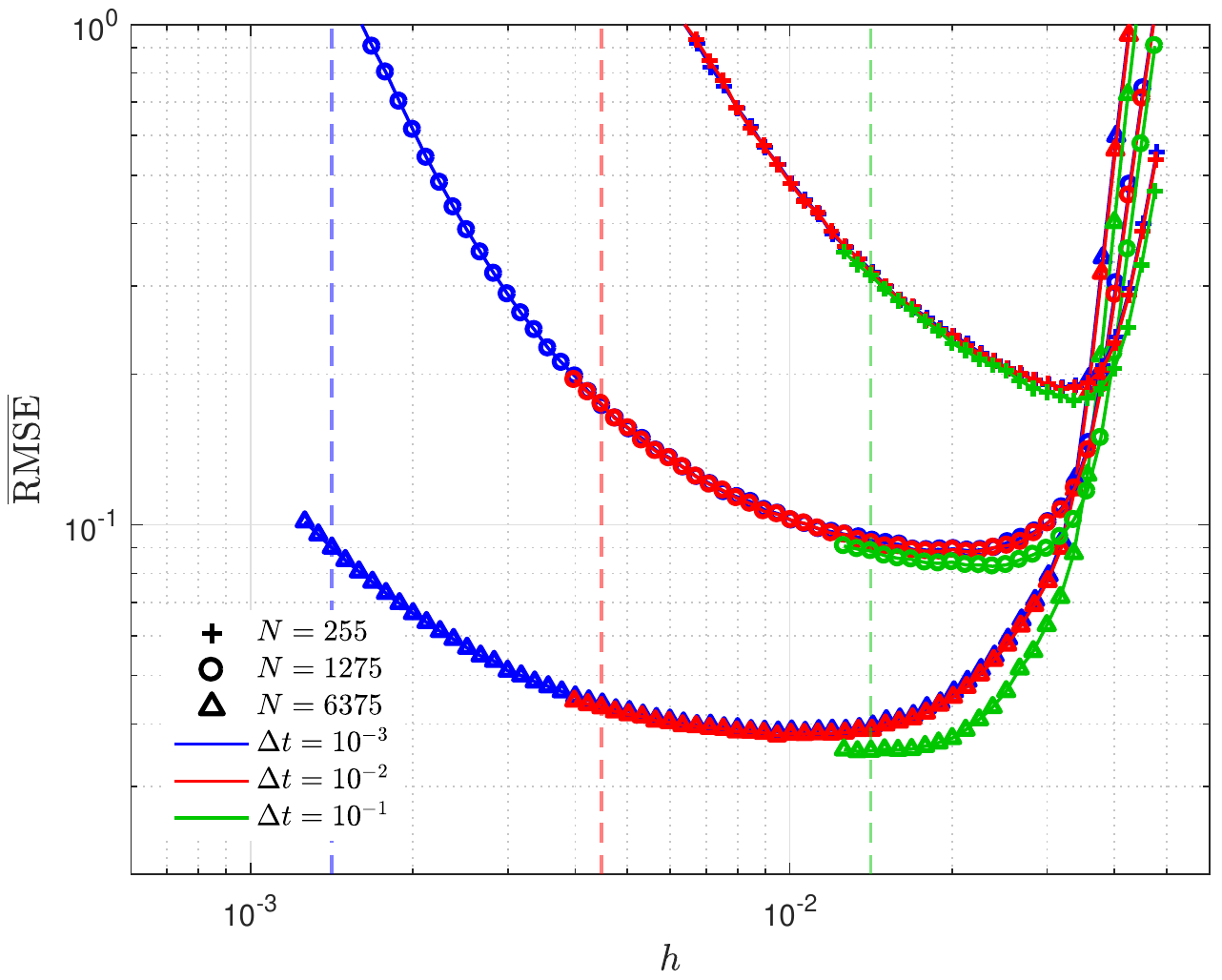}
    \caption{Numerical results for randomly-spaced, stationary particles. Averaged $\mathrm{RMSE}$ \eqref{RMSE}, as a function of the kernel bandwidth $h$, is given for different combinations of $N$ and $\Delta t$. The dashed, semitransparent vertical lines indicate the values of $h$ that correspond to $\beta=1$ ($h = \ell = \sqrt{2D\dt}$) for each value of $\Delta t$.}
    \label{fig:RMSE_rand}
\end{figure}

\begin{figure}[t]%
    \centering
    \includegraphics[width=1\textwidth]{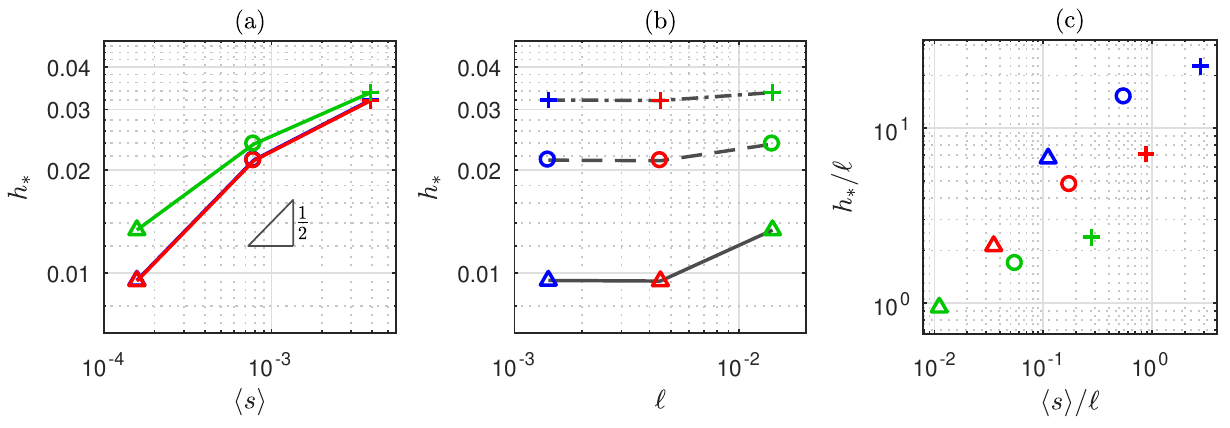}
    \caption{Numerical results for randomly-spaced, stationary particles. $\left( \mathrm a \right)$ Bandwidth $h_*$ associated with the minimum $\mathrm{RMSE}$ plotted against the average particle spacing $\langle s \rangle =L/N$, for different $\Delta t$ values (see color legend on Figure \ref{fig:RMSE_rand}). $\left( \mathrm b \right)$ Bandwidth $h_*$ associated with the minimum $\mathrm{RMSE}$ plotted against the dispersion distance $\ell=\sqrt{2D\Delta t}$ given different $N$ values (see marker legend on Figure \ref{fig:RMSE_rand}). $\left( \mathrm c \right)$ Lowest-error bandwidth $h_*$ against average particle spacing $\langle s \rangle =L/N$, both normalized by the dispersion distance $\ell=\sqrt{2D\Delta t}$.}
    \label{fig:hopt_rand}
\end{figure}

\begin{figure}[t]%
    \centering
    \includegraphics[width=1\textwidth]{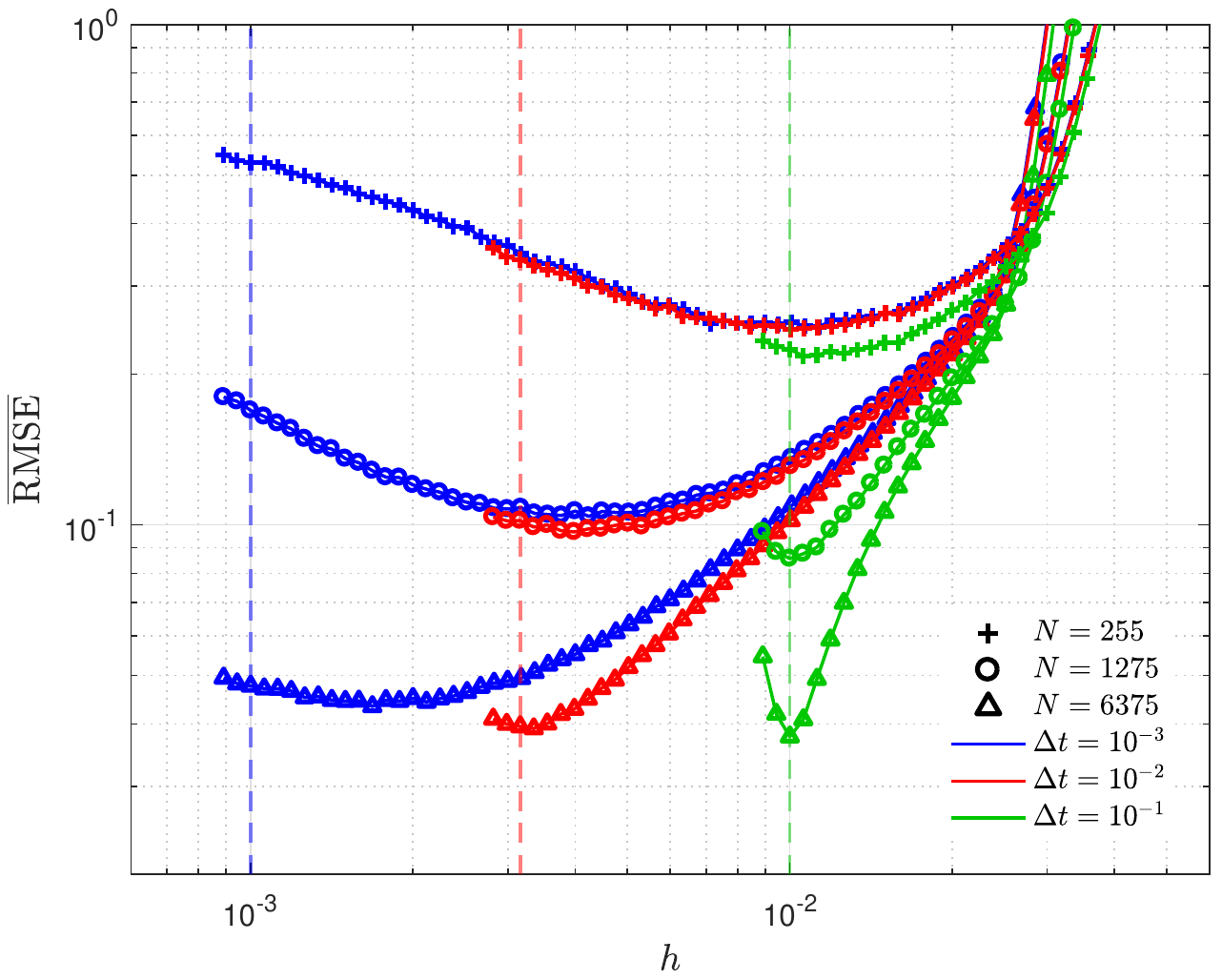}
    \caption{Numerical results for random-walking particles. Averaged $\mathrm{RMSE}$ \eqref{RMSE}, as a function of the kernel bandwidth $h$, is given for different combinations of $N$ and $\Delta t$. The dashed, semitransparent vertical lines indicate the values of $h$ that correspond to $\beta=1$ for each value of $\Delta t$.}
    \label{fig:RMSE_rwmt}
\end{figure}

\begin{figure}[t]%
    \centering
    \includegraphics[width=1\textwidth]{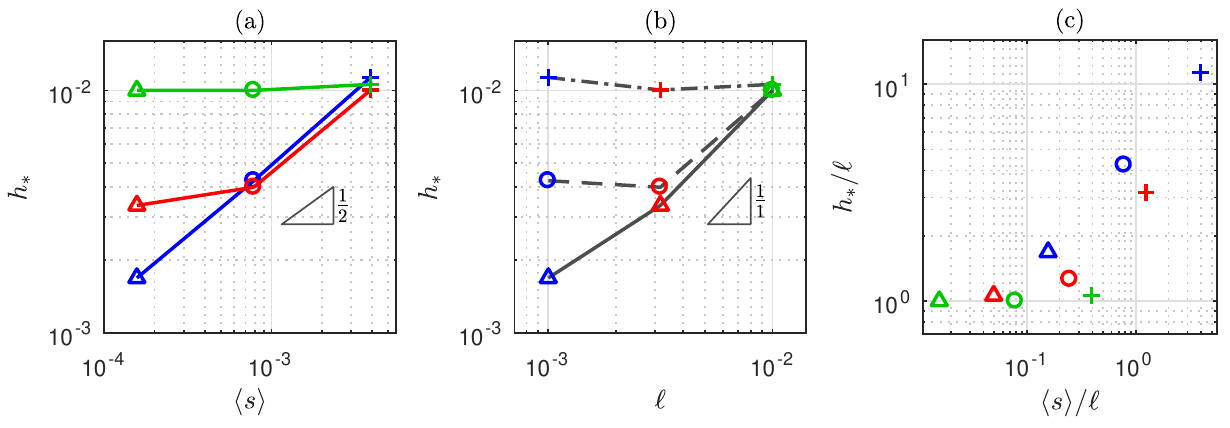}
    \caption{Numerical results for random-walking particles. $\left( \mathrm a \right)$ Bandwidth $h_*$ associated with the minimum $\mathrm{RMSE}$ plotted against the average particle spacing $\langle s \rangle =L/N$, for different $\Delta t$ values (see color legend on Figure \ref{fig:RMSE_rand}). $\left( \mathrm b \right)$ Bandwidth $h_*$ associated with the minimum $\mathrm{RMSE}$ plotted against the dispersion distance $\ell=\sqrt{2D_{\text{MT}}\Delta t}$ given different $N$ values (see marker legend on Figure \ref{fig:RMSE_rand}). $\left( \mathrm c \right)$ Lowest-error bandwidth $h_*$ against average particle spacing $\langle s \rangle =L/N$, both normalized by the dispersion distance $\ell=\sqrt{2D_{\text{MT}}\Delta t}$.}
    \label{fig:hopt_rwmt}
\end{figure}

To analyze the effect of the kernel bandwidth $h$ on SPH transport simulations, we study a simple case of 1D constant dispersion, where the initial condition is a Dirac delta pulse located at the center of the domain, $x = 0.5$ [L].
\rblue{This initial condition is represented in the numerical model by placing a particle with unitary mass of the solute at the center of the domain.}
For simplicity, the model has no units.
The dispersion coefficient is fixed as $D=10^{-3}$ [L$^2$T$^{-1}$], and the total simulation time is $T=4$ [T].
The analytical solution is then a Gaussian with variance $\sigma^2=2DT$ (see Figure \ref{fig:example}), or, to be more precise, the analytical solution is a normalized, $N$-bin histogram populated with evaluations of the density of a Normal distribution, $\cN(0.5, \sigma^2)$, at the positions of the particles.
We compare this analytical solution to the numerical results for a range of values of $h$, $N$ and $\Delta t$, using root-mean-squared error (RMSE) as the error metric, which is defined to be
\begin{equation}\label{RMSE}
    \mathop{\mathrm{RMSE}}\left(\vec {\ablue C}^{\mathrm{si}}\right)=\sqrt{\frac 1N \sum_{i=1}^N{\left({\ablue C}_i^{\mathrm{si}}\left(T\right)-{\ablue C}_i^{\mathrm{an}}\left(T\right)\right)^2}},
\end{equation}
where $\vec {\ablue C}^{\mathrm{an}}\left(T\right)$ is the analytical solution vector at time $T$, $\vec {\ablue C}^{\mathrm{si}}\left(T\right)$ is the corresponding result from a given simulation.\rblue{, and $\langle s\rangle \defeq L/N$ is the expected inter-particle spacing for $N$ particles and a domain of length $L$.
The normalization by $\langle s\rangle$ allows us to compare simulations with different particle numbers, as the $\mathrm{RMSE}$ for these normalized errors in particle mass is made equivalent to a standard error for the concentrations.}

For our numerical experiments, the $N$ particles are initially distributed over a fixed interval $\left[0,L\right]$, with $L=1$ [L].
\ablue The Dirac delta initial condition is represented in the numerical model by placing a particle with concentration $N/L$ at the center of the domain. \aend
We compare three different types of simulations: equally-spaced, stationary particles (Section \ref{sub:equally_spaced_stationary_particles}), randomly-spaced, stationary particles (Section \ref{sub:randomly_spaced_stationary_particles}), and particles moving by Brownian motion random walks (Section \ref{sub:random_walking_particles}).
For the latter two cases, initial particle positions are assigned according to draws from a uniform, $\cU(0, 1)$, distribution, and ensembles of $9\,520$ and {\agray $1\,660$} realizations of each configuration, respectively, are performed in order to obtain a smooth estimation of the expected error by averaging over the ensemble.
For fixed values of $N$ and $\Delta t$, we define $h_*$ as the bandwidth for which the lowest average $\mathrm{RMSE}$ is obtained, i.e.,
\begin{equation}\label{hopt}
    h_*=\mathop{\mathrm{argmin}}_{h > 0}\left(\bar{\mathop{\mathrm{RMSE}}\left(\vec {\ablue C}^\mathrm{si}; h\right)}\right),
\end{equation}
where $\bar{\mathop{\mathrm{RMSE}}\left(\vec {\ablue C}^\mathrm{si}; h\right)}$ is the average $\mathrm{RMSE}$ over all realizations.
\subsection{Equally-spaced, stationary particles}
\label{sub:equally_spaced_stationary_particles}

Figure \ref{fig:RMSE_equi} shows $\mathrm{RMSE}$ \eqref{RMSE} as a function of $h$ for different values of $N$ and $\Delta t$, for simulations with evenly-spaced, stationary particles.
In this case we observe a high degree of overlap between the curves, since marginal changes in $N$ and/or $\Delta t$ do not always have a significant effect on the simulation results.
The simple explanation for this is that, for a fixed $\dt$ that implies a given dispersion distance, $\ell = \sqrt{2 D \dt}$, increasing $N$ beyond a certain point does nothing to improve the ``resolution'' of the simulation, and the reverse also holds.
We see that, given a high enough density of particles ($N$ sufficiently large), the closest possible representation of the dispersion equation (lowest $\mathrm{RMSE}$) occurs for $\beta=1$.
\ablue In other words, for evenly-spaced particles, the smoothing kernel associated with $\beta=1$ is virtually free of numerical error when used in the SPH algorithm, as it in fact matches the analytical solution of the solute's dispersion over a time step. \aend
It is worth noting here that this value of $\beta=1$ does not correspond to the particle co-location algorithm, given in \eqref{arb_rxn_MT} \cite[see][]{Benson_arbitrary}, but to the generalization of the Green's function algorithm instead \cite{mass_trans_acc}, which is discussed in Section \ref{sec:link}.
From a physical point of view, using a kernel bandwidth larger than $\ell$ ($\beta<1$), could be seen as equivalent to assuming that the solute mass carried by each particle is Gaussian-distributed in space over some support, rather than a Dirac delta, prior to the start of the time step \cite{schmidt2017}.
This is consistent with the fact that, for low $N$, the $\mathrm{RMSE}$ can be reduced (up to a certain point) by using a larger kernel; i.e., the assumption that each particle is distributed over some support can mitigate the need for more particles.
Conversely, choosing a kernel bandwidth significantly smaller than $\ell$ $(\beta>1$), in addition to not having a clear physical meaning, generates numerical instabilities because the mass transfer between two particles in one time step may be larger than the difference between their masses (see \eqref{s1e7}).
As a result, these cases are excluded from the results shown in Figure \ref{fig:RMSE_equi}.

Some of the aforementioned relations can be better observed in Figure \ref{fig:hopt_equi}.
Given a coarse time discretization (Figure \ref{fig:hopt_equi}$\left( \mathrm a \right)$, green curves and markers), $h_*$ does not depend on $s$, and $h_*=\ell$.
Given a finer time discretization and a low particle density, we have the relation $h_* \propto s$ (see the linear trend, for large $s$, in the yellow curves of Figure \ref{fig:hopt_equi}$\left( \mathrm a \right)$). {\aviolet This proportionality is consistent with the known theoretical behavior for the truncation error of the SPH interpolation, given evenly-spaced particles \cite{Quinlan_TruncationError}.}
In examining the relation of $h_*$ to the dispersion distance $\ell=\sqrt{2D\Delta t}$ in Figure \ref{fig:hopt_equi}$\left( \mathrm b \right)$, we observe that, for sufficiently high values of $N$ and $\Delta t$, we have $h_*=\ell$ (corresponding to $\beta=1$, see the clearly distinguished minima in Figure \ref{fig:RMSE_equi}), and otherwise, $h_*\simeq s/\sqrt{2}$ (the curves with less pronounced minima in Figure \ref{fig:RMSE_equi}).
All these relations are summarized by the two distinguishable regimes that can be seen in Figure \ref{fig:hopt_equi}$\left( \mathrm c \right)$, wherein $h_*$ and $s$ are non-dimensionalized via scaling by the dispersion distance $\ell$.

\subsection{Randomly-spaced, stationary particles} 
\label{sub:randomly_spaced_stationary_particles}

The numerical results for randomly-distributed particles show less distinct trends, in terms of matching the analytical solution, than those seen for the evenly-distributed particles of Section \ref{sub:equally_spaced_stationary_particles}, and this can be seen in Figure \ref{fig:RMSE_rand}.
In this case, the $\mathrm{RMSE}$ does not always have such a clearly identifiable minimum in the vicinity of $h_*$, nor does $h_*$ reliably correspond to $\beta = 1$, as we saw in Section \ref{sub:equally_spaced_stationary_particles}.
\aviolet Rather, its behavior appears to roughly agree with the theoretical SPH truncation error for randomly-spaced particles \cite{Tartakovsky2016,Quinlan_TruncationError}, which can be expressed as the summation of two terms: the smoothing error, which scales with $h$; and the quadrature error, which scales with $\langle s\rangle /h$, where $\langle s\rangle$ is the expected particle separation (here, $\langle s\rangle=L/N$). Balancing these two terms results in $h_*\propto\sqrt{\langle s\rangle}$, and hence for that choice of bandwidth the truncation error scales with $\sqrt{\langle s\rangle}$. This is consistent with the results shown in Figure \ref{fig:RMSE_rand}, where \aend \rviolet{We note that}, given $h=h_*$ {\aviolet (i.e., considering only each curve's minimum)}, the $\mathrm{RMSE}$ scales with the particle number as $\mathrm{RMSE} \propto N^{-1/2}$. \rviolet{This kind of scaling is characteristic of the noise of a density estimator with respect to the number of data points. This could indicate that, by choosing $h=h_*$, the only error introduced by reducing the particle density is the natural extra noise associated with a lower density of information.}

It is only when $\Delta t$ adopts large values that it appears to have a noticeable influence on the $\mathrm{RMSE}$\rred{, which seems to stabilize for small enough $\Delta t$ (as it can be seen from the nearly coincident red and blue lines and markers in Figure . As a result, for small $\Delta t$, $h_*$ is mainly a function of the expected particle separation $\langle s \rangle$, or equivalently, the particle number $N$. As $N$ increases, $h_*$ becomes smaller and approaches $\ell$. In addition, for increasing $N$, the $\mathrm{RMSE}$ becomes flatter (less sensitive to changes in $h$)}.
This behavior is also evident in the relative insensitivity of $h_*$ to $\ell$, as can be seen in Figure \ref{fig:hopt_rand}$\left( \mathrm b \right)$.
\rred{However, }In Figure \ref{fig:hopt_rand}$\left( \mathrm a \right)$ we see that the relation of $h_*$ to the average particle spacing $\langle s \rangle$ is not linear{\agray, not even for small $\Delta t$}, unlike in the evenly-spaced particle case. Instead, we observe a range of slopes in the log-log space (about $1/2$ and lower), which can be related to the aforementioned truncation error \cite{Quinlan_TruncationError}, which is minimized when $h\propto\sqrt{\langle s \rangle}$\rred{suggesting that a direct proportionality does not exist, i.e., other factors, including $\ell$, may also be affecting the value of $h_*$, especially for low $\langle s\rangle$ (high particle density)}.
Unlike the equally-spaced case (Figure \ref{fig:hopt_equi}$\left( \mathrm c \right)$), we do not observe a single linear trend in Figure \ref{fig:hopt_rand}$\left( \mathrm c \right)$ for the relationship between $h_*/\ell$ and $\langle s \rangle/\ell$.
Rather, we observe the general tendency that  $\langle s \rangle \to 0$ implies $h_* \to \ell$.
For the range of tested values, a relatively high particle density, of $\langle s\rangle\lesssim0.01\ell$, is required to observe the relation $h_*\simeq\ell$.

\subsection{Random-walking particles} 
\label{sub:random_walking_particles}


The same set of simulations are also conducted for a hybrid model in which the dispersion coefficient is partitioned as
\begin{equation}\label{splitD}
    D=D_{\text{RW}}+D_{\text{MT}},
\end{equation}
where $D_{\text{MT}}$ is the dispersion coefficient used in the SPH/MTPT algorithm described in the previous section, and particles move by Brownian motion, according to the Langevin equation. For a time discretization $\{ t_1,t_2,...,t_n\}$, with $t_{k+1}=t_k+\Delta t$,
\begin{equation}\label{RW}
    X_i^{k+1}=X_i^k+\xi_i^k \sqrt{2D_{\text{RW}}\Delta t},
\end{equation}
where $X_i^k\defeq X_i(t_k)$, and $\xi_i^k$ is a random number drawn from a standard normal, $\cN(0, 1)$, distribution.
With an appropriate choice of $D_{\text{RW}}$ and $D_{\text{MT}}$, this type of approach can be used to give a separate treatment to the non-mixed spreading (RW) and the actual mixing (MT).
Several authors \cite{Gelhar1979,Gelhar_1983,Cirpka_1999,Werth_focus} have suggested that these correspond to the anisotropic spreading (longitudinal minus transverse hydrodynamic dispersion) and the isotropic mixing (molecular diffusion plus transverse hydrodynamic dispersion) parts of the dispersion tensor, respectively.
Here we simply set $D_{\text{RW}}=D_{\text{MT}}=D/2$.
Note that, for this partitioning, random walks do not significantly perturb spatial concentrations about their expected value.\rviolet{(the analytical, well-mixed solution); i.e., random walks do not introduce excess noise to the simulated solution.} {\aviolet That is, the concentration difference between two spatially coincident particles is negligible, meaning that the concentrations at a given time vary ``smoothly'' with the particle positions $X_i$ (see Figure \ref{fig:example}, yellow markers)}. 
This is because particles exchange mass at the same rate at which they diffuse by Brownian motion.
For this reason, we can study the influence of $h$ on the numerical results when particles are random-walking and compare to the case where particles are stationary (as in Sections \ref{sub:equally_spaced_stationary_particles} and \ref{sub:randomly_spaced_stationary_particles}), without introducing the concentration variance that would be otherwise (purposefully) induced by setting $D_{\text{RW}}\gg D_{\text{MT}}$.
{\aviolet Since, at $t=0$, there is only one particle with nonzero concentration}, a strong variability in the results is introduced by the random motion of that particle in the initial stages of the simulation, when it is carrying nearly all the {\agray solute} mass in the system.
For this reason, in order to favor faster convergence of the $\mathrm{RMSE}$ with the number of simulations, we set that singular particle to be motionless and to use the full dispersion coefficient in its mass-transfer calculations (i.e., for that particle, $D_{\text{MT}} = D$ and $D_{\text{RW}} = 0$).
An alternative approach to overcome the same issue could be to use more particles to represent the initial Dirac delta condition.

The behavior of the $\mathrm{RMSE}$ in this case (Figure \ref{fig:RMSE_rwmt}) can be seen as occupying a middle ground between the equally-spaced (Figure \ref{fig:RMSE_equi}) and the randomly-spaced (Figure \ref{fig:RMSE_rand}), stationary cases.
The distribution of particle spacings in the random-walking case at any given time is identical to the stationary randomly-distributed case, but in the former, the expected, or time-averaged, particle spacing distribution is much narrower, approximating the stationary, evenly-spaced case in that sense.
For that reason, we do expect the value of $h_*$ for a random-walking model, in the context of this specific example, to be bounded between the two extreme stationary cases, which may be thought of as the most ordered and disordered systems, respectively.
Note, however, that the actual values of the $\mathrm{RMSE}$ in Figure \ref{fig:RMSE_rwmt} are on the same order of magnitude as for the randomly-distributed, stationary particles (Figure \ref{fig:RMSE_rand}), and they can be even higher.
This may be attributed to the added natural variability of Brownian random walks used to represent half of the dispersion, as opposed to the deterministic nature of mass transfers.
For high enough $N$ and $\Delta t$, we can see that $\mathrm{RMSE}$ minima occur at $h_*=\ell$ and are strongly pronounced. Otherwise, we see milder minima and $h_*>\ell$, similarly to what is observed for equally-spaced particles (Figure \ref{fig:RMSE_equi}).
In these regions of milder minima, we see the approximate scaling $\mathrm{RMSE} \propto N^{-1/2}$ given $h=h_*$, which, in this behavior, is similar to the randomly-spaced, stationary case (Figure \ref{fig:RMSE_rand}).

\rred{Once again, we do not see a linear scaling of $h_*$ with the average spacing $\langle s\rangle$ in Figure $\left( \mathrm a \right)$, but instead an apparent relation of $h_*\propto \sqrt{\langle s \rangle}$. As previously mentioned, a possible explanation for this is that other factors, such as the dispersion distance, $\ell$, are affecting the value of $h_*$.}
We see that for a fine time discretization (blue line in Figure \ref{fig:hopt_rwmt}$\left( \mathrm a \right)$), we have $h_*\propto \sqrt{\langle s \rangle}$, {\aviolet which, as mentioned in Section \ref{sub:randomly_spaced_stationary_particles}, indicates that $h_*$ in these regimes is mainly controlled by the truncation error of the spatial interpolation}\rred{approximately}.
On the other hand, we see a clear trend that $h_*=\ell$ for large enough $N$ and $\dt$, as evidenced by the triangle symbols and green markers in Figure \ref{fig:hopt_rwmt}$\left( \mathrm b \right)$.
As in the previous cases, $h_*$ departs from $\ell$ at some threshold as the relative spacing $\langle s\rangle/\ell$ increases. Like in the stationary, randomly-spaced case, and unlike the equally-spaced case, this threshold value for $\langle s\rangle/\ell$ appears to depend on $\ell$ (i.e., no single linear trend is observed in Figure \ref{fig:hopt_rwmt}$\left( \mathrm c \right)$, unlike in Figure \ref{fig:hopt_equi}$\left( \mathrm c \right)$).
Nevertheless, for the range of tested values, $h_*\simeq\ell$ for $\langle s\rangle\lesssim0.1\ell$.


\section{Summary and discussion} 
\label{sec:summary_and_discussion}

In this paper, we demonstrate an equivalence between the Lagrangian SPH (smoothed particle hydrodynamics) and MTPT (mass transfer particle tracking) methods for simulating dispersion, provided that the spatial kernel being employed is Gaussian.
These two methods originate from completely different interpretations.
The SPH community views their methods (classically speaking, as recent work has included random walks in SPH simulations \cite{herrera_2017}) as solving the dispersion equation by projecting the particles onto the continuum using radial basis functions (kernels) and approximating the solution on that kernel space.
The random walk particle tracking community views the MTPT methods considered in this paper in two ways: (i) a first-principles approach, wherein mass-transfers between moving particles are scaled by the probability that these particles co-locate via dispersion; (ii) a discretization of the Green's function for the dispersion equation, in which a particle's solute mass is spread in space via mass-transfers to its nearest neighbors.
Previously, these two MTPT methods were considered to be distinct approaches, and neither had rigorous proofs associated with it.
As a result of this work, however, both of these MTPT methods now inherit a rigorous theoretical underpinning from the SPH literature.

The numerical investigations we conduct yield compelling results regarding the proper Gaussian kernel bandwidth for particle tracking simulations.
We see strong evidence that a kernel with bandwidth $h = \ell = \sqrt{2D\dt}$ (i.e., imposing $\beta=1$) is the ideal choice, provided there is a ``dense enough'' spatial distribution of particles.
This makes intuitive/physical sense because, with bandwidth $\ell$, this Gaussian function is the fundamental solution of the dispersion equation.
In other words, aside from the error introduced in the normalization step, using this kernel for mass transfer is not an approximation, but rather a semi-analytical solution of the dispersion in a time-step of length $\dt$.
We also observe that, counter-intuitively, a coarser time-discretization may be a better choice than a finer one, if that allows one to use bandwidth $\ell$.
However, there may be cases in which the intent is to reproduce the dispersion equation without the distortion associated with a low particle density (a subject that we discuss below), but a high particle density cannot be afforded, computationally (as may be likely to occur in multi-dimensional systems).
If, in these cases, the use of a long time-step would generate other forms of error (for instance, in the chemical reactions), then a wider kernel bandwidth than $\ell$ {\agray (following the traditional SPH bandwidth selection rules-of-thumb)} may be a better choice when seeking a compromise between accuracy and efficiency.
One way to think of this is to consider the wider-bandwidth particle to be a ``macro-particle,'' or cluster of smaller particles, that is distributed in space over some support volume.

Additional conclusions can be drawn from each of the individual cases tested in Section \ref{sec:numerical}.
In the equally-spaced, stationary particle case, $h = \ell$ is clearly the optimal bandwidth choice, provided that $N$ is sufficiently large, as to capture the magnitude of dispersion, described by $\ell = \sqrt{2D\dt}$ (i.e., particles must be close enough to ``see'' one another).
\rred{This is because the equally-spaced particle condition represents the closest $N$-particle approximation of a perfectly-mixed system, so the kernel/bandwidth that has the lowest $\mathrm{RMSE}$, as compared to the analytic, well-mixed solution, is clearly optimal for that problem.}

Considering the randomly-distributed, stationary particle case, we see a different story, in that $\mathrm{RMSE}$ tends to be more related to average inter-particle spacing, $\langle s \rangle = L / N$, than it is to the dispersion distance, $\ell$. {\ared This is most likely because, for the range of $N$ and $\Delta t$ values tested, the $\mathrm{RMSE}$ is dominated by the truncation error of the SPH interpolation. Nevertheless,} according to some authors in particle methods \cite[e.g.,][]{Paster_JCP, Ding_WRR}, the distortion of the numerical solution caused by heterogeneity in the inter-particle spacing and low particle densities can represent incomplete mixing conditions, rather than being just a numerical error.
If we subscribe to this view, then the randomly-spaced case represents areas in which particles are poorly-mixed and remain poorly mixed for the duration of the simulation.
From that perspective, using the $\ell$ bandwidth would only be capturing the ``average mixedness'' of such a simulation, fully simulating diffusive mixing in well-mixed areas and under-simulating mixing in poorly mixed areas.
In light of this, the increase in $\mathrm{RMSE}$ could be thought of not as an error, but as desirable deviations from the well-mixed solution, due to physically meaningful areas of poor mixing.

For the case of random-walking particles, we find that the qualitative behavior of the $\mathrm{RMSE}$ with respect to the bandwidth $h$ can be placed in a middle ground between the other two scenarios.
In fact, the minima ($h_*$) are found to be bounded in this case between the two former cases. {\agray It is clear from the results that, despite the particle disorder, the dependence of the $\mathrm{RMSE}$ on $h$ should not be understood as a function of the particle density alone. Instead, the error originated in deviating from the dispersion kernel bandwidth $h=\ell$ should also be considered.}
Again, if the effects of particle disorder on the numerical solution are considered to be physically meaningful, it makes sense that random-walking particles are closer to representing a well-mixed system (distinguishable by $h_*=\ell$) than stationary randomly-distributed particles, since in this case the poorly-mixed areas are not persistent in time.

We believe the results of our numerical experiments are relevant in a general sense, despite representing the specific simple case of a Dirac delta initial condition in a one-dimensional setting.
This particular dispersion problem, where one initial concentration pulse spreads by dispersion, is no doubt the simplest one; however, any more complex problem can be thought of as unions of Dirac delta initial conditions, at least from a computational/discrete standpoint.
As long as the physics are being captured on a local, particle level, as is demonstrated here, more complicated conditions will also be properly simulated.
Additionally, we expect the scaling with $s$ and $\ell$ to be analogous {\aviolet for isotropic dispersion} in higher dimensions because mass transfers are merely a function of Euclidean distance between particles, and hence not substantively different in higher spatial dimensions. 
However, the scaling relations will likely need to be reformulated in terms of fill distance, rather than the simple inter-particle spacing we see here in 1D. \aviolet Besides, the analysis performed in Section \ref{sec:numerical} would undoubtedly become more complex in the case of anisotropic and spatially variable dispersion.\aend

The traditional SPH extension to anisotropic dispersion entails a more complicated expression for $\hat D_{ij}$ in \eqref{s1e1}, while maintaining the isotropy of the kernel $W$, and this approach may result in negative concentrations \cite{herrera_2013}. This is in contrast to the more straightforward extension of traditional MTPT to anisotropic dispersion, which would involve redefining $W$ as an anisotropic multi-Gaussian with variance $2\Delta t \cdot g\left(\vec{D}\left(\vec X_i\right),\vec{D}\left(\vec X_j\right)\right)/\beta$, where $g$ is some averaging function. The subject of anisotropy is out of the scope of this paper and should be addressed in future work. Nevertheless, as mentioned in Section \ref{sec:numerical}, another suitable approach to reproducing anisotropic dispersion would be to split the dispersion tensor between an isotropic and an anisotropic part, using the isotropic SPH/MTPT method addressed here to simulate the former and reproducing the latter with random walks.

Open questions do remain in this area.
For instance, we only consider the Gaussian kernel in our analysis and results.
Other kernels are commonly used in the SPH literature, and compactly-supported kernels are known to result in computational speedup.
A standard choice is the compactly-supported Wendland kernel that has been shown to approach a Gaussian in the infinitely-smooth, limiting case \cite{wendland_gaussian}.
How much error is introduced by this approximation, and how does this compare to the common practice or imposing a cutoff distance of $3h$ for mass transfers, as is commonly done in the particle tracking literature?

The hybridization of SPH/MTPT with random walks is a very recent technique that, to date, has not been studied in depth.
In this work, we compare the numerical results from one such model with an analytical solution in the particular case wherein the simulation of the full dispersion tensor is partitioned equally between random walks and mass transfers.
If the purpose of this hybridization is to simulate a two-scale system {\ared (as in \cite{herrera_2017})} in which the random walk accounts for spreading and the mass transfer accounts for mixing, it would be proper for the magnitude of mixing to be much smaller than that of spreading, in order to generate states of local disequilibrium (as, for instance, to simulate the effect of local heterogeneities in porous media).
Hence, further investigation is needed in this area, in order to: (i) analyze the effect of using different spreading/mixing ratios, and (ii) evaluate the capability of this kind of model to correctly reproduce the generation, propagation, and decay of sub-scale concentration variance.



\bibliography{SPH_equiv_AWR}

\end{document}